\documentclass[runningheads]{llncs}
\usepackage[T1]{fontenc}
\usepackage{makecell}
\usepackage{graphicx}
\usepackage{url}
\usepackage{multirow}
\usepackage{amssymb}
\usepackage{cite}
\usepackage{bbding}
\usepackage{caption}
\usepackage{bbding}
\usepackage[colorlinks, linkcolor=blue, filecolor=blue, filecolor=blue, urlcolor=blue, citecolor=blue]{hyperref}
\usepackage{appendix}
\usepackage{color}
\begin{document}
\title{Uni4Eye: Unified 2D and 3D Self-supervised Pre-training via Masked Image Modeling Transformer for Ophthalmic Image Classification}

% \thanks{}
%
\titlerunning{Uni4Eye: Unified 2D and 3D Ophthalmic Image Pre-training}
\author{
Zhiyuan Cai\inst{1,1,\footnotemark[1]} \and
Li Lin\inst{1,2,\footnotemark[1]} \and
Huaqing He\inst{1}\and
Xiaoying Tang\textsuperscript{1(\Envelope)}}
%
% \institute{Anonymous Organization
% }
\authorrunning{Cai et al.}
\institute{Department of Electronic and Electrical Engineering,\\
Southern University of Science and Technology, Shenzhen, China\\
\email{tangxy@sustech.edu.cn}\\\and
Department of Electrical and Electronic Engineering,\\
The University of Hong Kong, Hong Kong SAR, China}
\maketitle              % typeset the header of the contribution
\renewcommand{\thefootnote}{\fnsymbol{footnote}}
\footnotetext[1]{Z. Cai and L. Lin contributed equally to this work.}
\begin{abstract}
A large-scale labeled dataset is a key factor for the success of supervised deep learning in computer vision. However, a limited number of annotated data is very common, especially in ophthalmic image analysis, since manual annotation is time-consuming and labor-intensive. Self-supervised learning (SSL) methods bring huge opportunities for better utilizing unlabeled data, as they do not need massive annotations. With an attempt to use as many as possible unlabeled ophthalmic images, it is necessary to break the dimension barrier, simultaneously making use of both 2D and 3D images. In this paper, we propose a universal self-supervised Transformer framework, named Uni4Eye, to discover the inherent image property and capture domain-specific feature embedding in ophthalmic images. Uni4Eye can serve as a global feature extractor, which builds its basis on a Masked Image Modeling task with a Vision Transformer (ViT) architecture. We employ a Unified Patch Embedding module to replace the origin patch embedding module in ViT for jointly processing both 2D and 3D input images. Besides, we design a dual-branch multitask decoder module to simultaneously perform two reconstruction tasks on the input image and its gradient map, delivering discriminative representations for better convergence. We evaluate the performance of our pre-trained Uni4Eye encoder by fine-tuning it on six downstream ophthalmic image classification tasks. The superiority of Uni4Eye is successfully established through comparisons to other state-of-the-art SSL pre-training methods.

\end{abstract}
%
%\cite{lee2021fundus, VIRMANI2019215}
%
\section{Introduction}

Recently, supervised deep learning methods have been found to perform comparably to human experts in various medical image analysis tasks such as disease classification \cite{li2021rotation} and structure segmentation \cite{lin2021bsdanet, JOINED}, benefiting from supervision of large-scale labeled datasets \cite{tan2020efficientnet}. However, manual delineation is time-consuming and labor-intensive, especially for large-scale datasets. Besides, fully supervised learning may somehow limit the model performance in some scenarios, such as in the Noisy Label \cite{cordeiro2021longremix} scenario.

To address these issues, self-supervised learning (SSL) methods have been gaining increasing research interest in the medical image analysis realm. SSL methods can be mainly categorized into generative and discriminative approaches \cite{atito2021sit}. For generative approaches, \cite{donahue2019large} models the distribution of the data based on a GAN \cite{goodfellow2014generative} framework, which is very computationally expensive. On the other hand, discriminative approaches focus on obtaining better generalized representations with relatively low computational burdens. Typically, discriminative approaches are implemented with contrastive learning frameworks \cite{huang2021lesionbased, he2020momentum, chen2020simple, cai2022corolla} or through novel pre-text tasks \cite{gidaris2018unsupervised, ye2019unsupervised}. The main shortcoming of contrastive learning methods is that they often focus on the main part of a medical image of interest but disregard contextual representations. Since the main parts are highly similar across different medical images, contrastive learning methods might fail, in which situation pre-text tasks accommodate better \cite{atito2021sit}. Recently, novel pre-text tasks have been explored, such as the Rubik's Cube Recovery task\cite{Cube} and the Masked Image Modeling (MIM) task \cite{mae, maskfeat}. MIM originates from the idea of masked signal modeling which refers to masking a portion of the input signals and trying to predict the masked signals. Lately, based on Vision Transformer (ViT) backbones, MIM attains huge success in SSL on natural image. For example, \cite{bao2021beit} and \cite{mae} employ MIM and get pre-trained on ImageNet-1k \cite{deng2009imagenet}, which respectively achieve 86.3\% and 87\% Top-1 accuracy.
%masked image modeling (MIM) \cite{chen2022context} task \cite{mae, maskfeat}, etc.

Nevertheless, the success of SSL has a prerequisite of massive datasets \cite{oliver2019realistic}. For instance, the recent success of Transformers on image classification \cite{dosovitskiy2021image} is mainly due to the large-scale ImageNet \cite{deng2009imagenet} dataset. However, for intelligent analyses of ophthalmic images, the sample sizes are usually very small. Ophthalmic image modalities can be categorized into 2D (e.g., fundus image \cite{lin2020sustech} and Fundus Fluorescein Angiography (FFA)) and 3D (e.g., Optical Coherence Tomography (OCT) and Optical Coherence Tomography Angiography (OCTA)). Because of the dimension barrier, current SSL approaches are typically designed for dimension-specific images \cite{chaitanya2020contrastive, CHEN2019101539, zhou2021preservational, taleb20203d}; that is, an SSL model can only accommodate either 2D or 3D images, which contradicts the intuitive motivation of employing as many as possible data for better performance.

% Several works have been attempted to address such problems.
% Several works have been attempted to address such problems. Masked Autoencoders (MAE) \cite{mae} is a famous work based on MIM task, which can be a solution to the small number of ophthalmic images. MAE is pre-trained on ImageNet-1k \cite{deng2009imagenet} but achieves over 87\% Top-1 accuracy, which outperforms all Vision Transformers (ViT) \cite{dosovitskiy2021image} variants in ImageNet-21k pre-training. Besides, to use as many unlabeled medical images as possible in SSL, it is necessary to break the dimension barrier (i.e., making it possible to jointly use both 2D and 3D images). \cite{xie2021unified} propose a Universal Self-supervised Transformer (USST) framework based on the student-teacher paradigm and contrastive learning to break the dimension barrier, which is effective but still disregards the learning of contextual representations.
 
In such context, we propose a simple yet effective framework that can learn universal representations from both 2D and 3D ophthalmic images, named \linebreak Uni4Eye. Uni4Eye is designed to perform dual MIM tasks with a ViT architecture. We design a two-branch switchable patch embedding layer in Uni4Eye to replace the origin patch embedding layer, which allows it to switch to different branches for patch embedding of 2D and 3D images. Furthermore, we employ a dual-branch decoder in our network and train it with different modeling/reconstruction tasks, so as to achieve more robust convergence and better representation. Additionally, we create so far the largest ophthalmic image dataset of multiple modalities as well as multiple dimensions, consisting a total of 95,978 samples. We name it as \emph{mmOphth}-v1, on which our proposed Uni4Eye gets pre-trained.

Collectively, our main contributions are three-fold: (1) To the best of our knowledge, this is the first time that a self-supervised pre-training framework is proposed to learn general visual representations of both 2D and 3D ophthalmic images. (2) We collect and create the largest ophthalmic image dataset of multiple modalities and of both 2D and 3D dimensions, named as \emph{mmOphth}-v1. This dataset will be made publicly available. (3) We conduct extensive experiments on six downstream classification tasks with four datasets involving common eye diseases. The superiority of our proposed Uni4Eye over other state-of-the-art (SOTA) self-supervised pre-training methods is successfully established on these tasks. The source code is available at \url{https://github.com/Davidczy/Uni4Eye}.

% For example, \cite{}, \cite{}, \cite{} and \cite{} have been proposed to construct image representations that are semantically meaningful from unlabelled data.
\vspace{-5pt}
\section{Methodology}
\vspace{-20pt}
\begin{figure}[htb]
    \centering\includegraphics[width=1\textwidth]{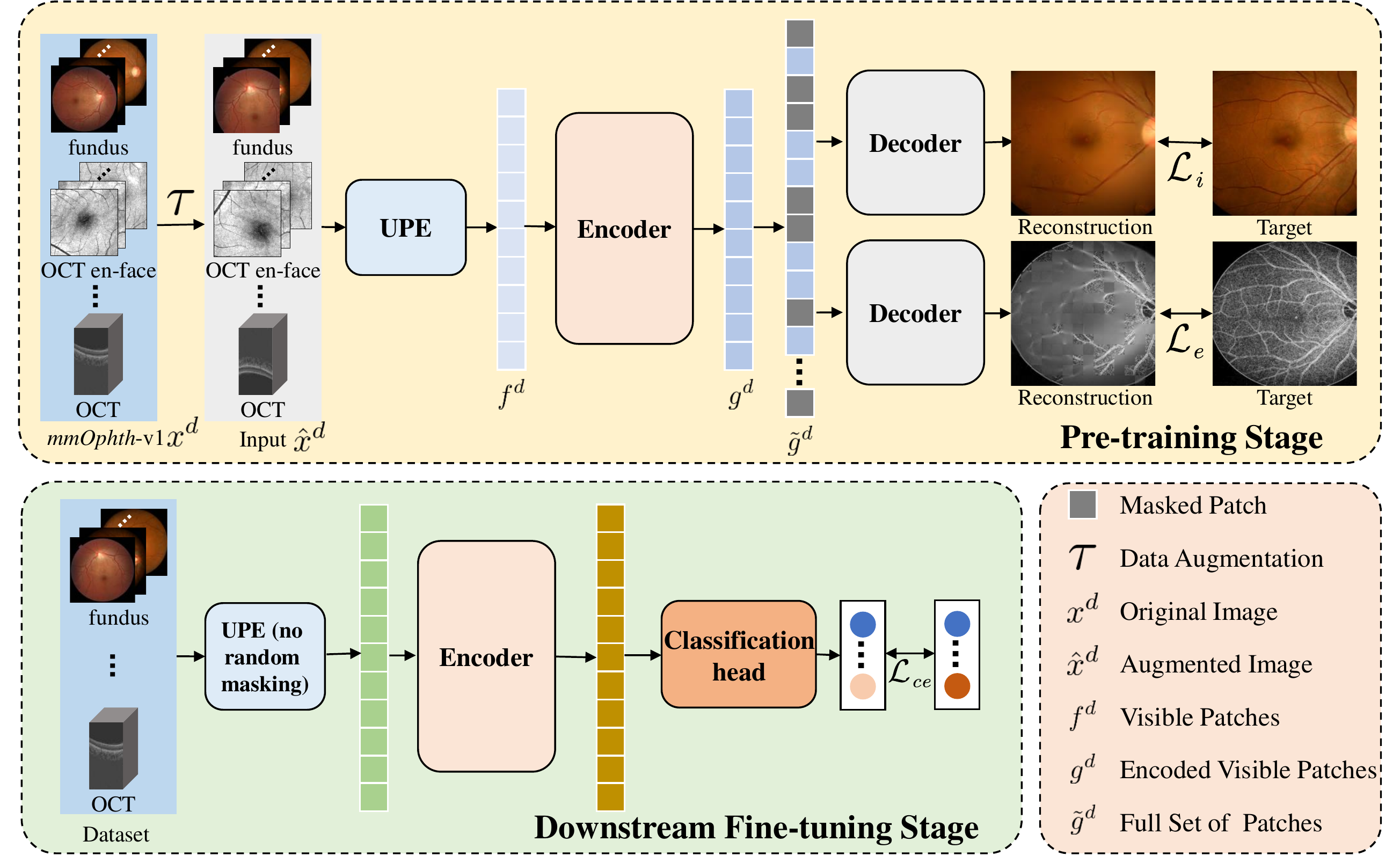}
     \caption{The overall framework of Uni4Eye.}
     \label{pipeline}
     \vspace{-0.5cm}
\end{figure}
The overview of our Uni4Eye is provided in Fig. \ref{pipeline}. There are three main components, including a Unified Patch Embedding (UPE) module, a ViT encoder and a dual-branch decoder. We first pre-train our encoder on two MIM self-supervised tasks in the pre-training stage and then fine-tune our model on different downstream tasks. As shown in Fig. \ref{pipeline}, the pre-training stage and downstream fine-tuning stage are respectively denoted as $P$ and $D$. Stage $P$ aims at training the encoder to generate more generalized and discriminative representations from different input ophthalmic images. Then, the UPE module and the ViT encoder in $D$ are utilized to load pre-trained parameters and continue to fine-tune on different downstream tasks to achieve better performance. For a downstream classification task, we adopt a fully-connected layer as the classification head to process features generated by the encoder and output prediction. We now delve into the details of UPE and the dual-branch decoder.
\vspace{-5pt}
\subsection{Unified Patch Embedding Module}
To make our self-supervised framework compatible with both 2D and 3D data and accommodate different downstream scenarios, we employ UPE as the patch embedding module. As shown in Fig. \ref{pipeline}, different images in the \emph{mmOphth}-v1 dataset can be directly fed into the UPE module, regardless of their dimensions. 
\vspace{-5pt}
{\setlength\abovedisplayskip{0.3cm}
\setlength\belowdisplayskip{0.3cm}
\begin{figure}[htb]
\vspace{-5pt}
\centering
     \includegraphics[width=0.75\textwidth]{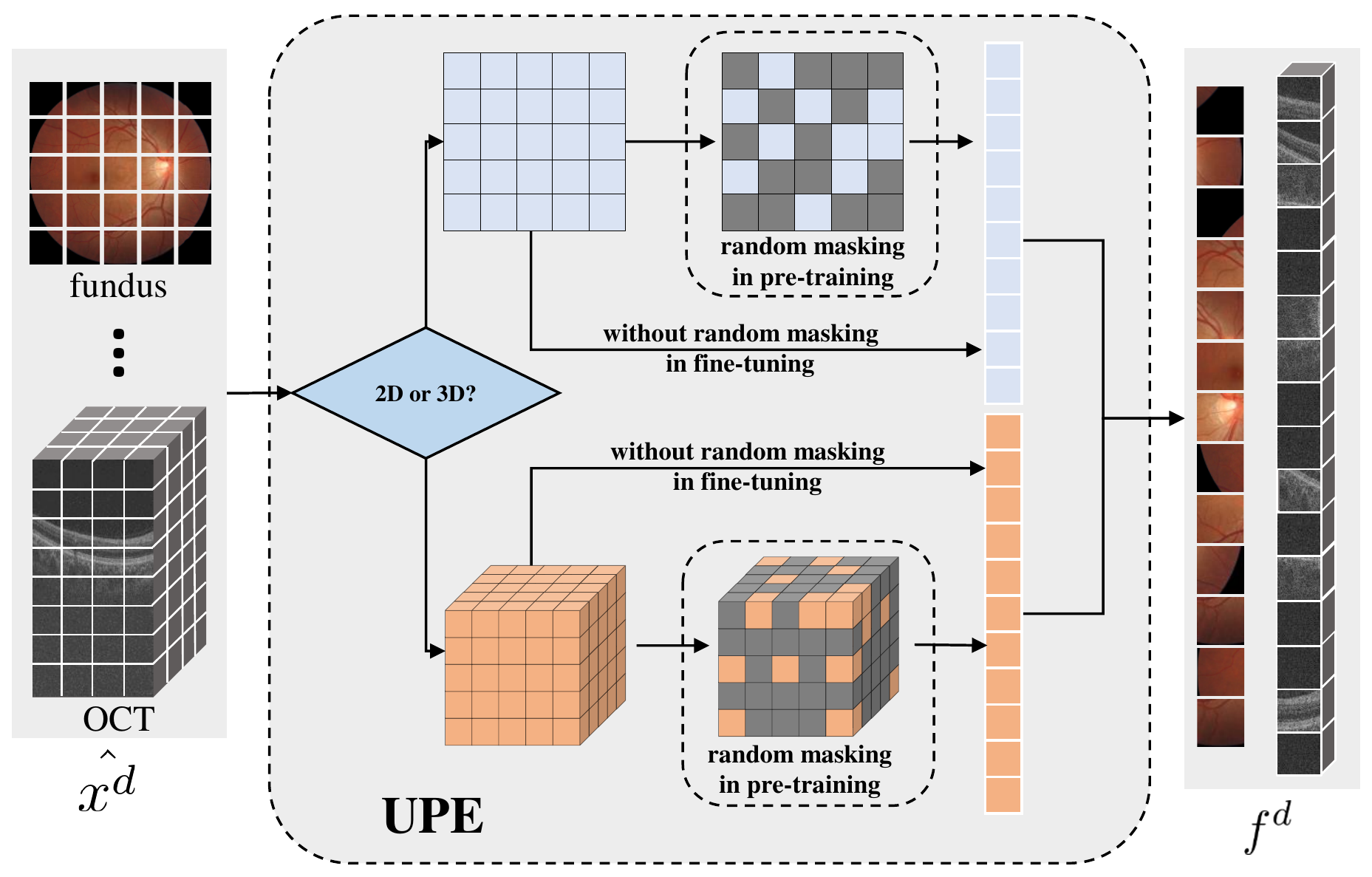}
     \caption{The structure of the Unified Patch Embedding module.}
     \label{UPE}
\end{figure}}
\vspace{-10pt}

Fig. \ref{UPE} illustrates the structure of UPE. Let an unlabeled training image sampled from \emph{mmOphth}-v1 be denoted by $x^{d}$, where $d \in \{2, 3\}$ represents the dimension of the image. Then, data augmentation $\tau$ is applied to $x^{d}$ to generate an input $\hat{x}^{d}$ of the UPE module. UPE switches $\hat{x}^{d}$ to specific patch embedding depending on the dimension of $\hat{x}^{d}$. Afterwards, a random masking strategy is employed to generate the masked patch embedding $f^d$ in stage $P$, while the strategy is skipped in stage $D$. To be more specific, since we divide an image into regular non-overlapping patches (2D square patches for 2D input and 3D cubic patches for 3D input), we follow a uniform distribution to sample random patches without replacement. Then, the remaining ones are masked out, which means these patches will not be fed into the encoder. Thus, the ViT encoder operates only on the visible patches but not the masked ones, which differs our proposed method from inpainting methods.

\vspace{-5pt}
\subsection{Dual-Decoder for Intensity and Edge Reconstruction}
{
\setlength\abovedisplayskip{0.3cm}
\setlength\belowdisplayskip{0.1cm}
\begin{figure}[htb]
\centering
     \includegraphics[width=0.65\textwidth]{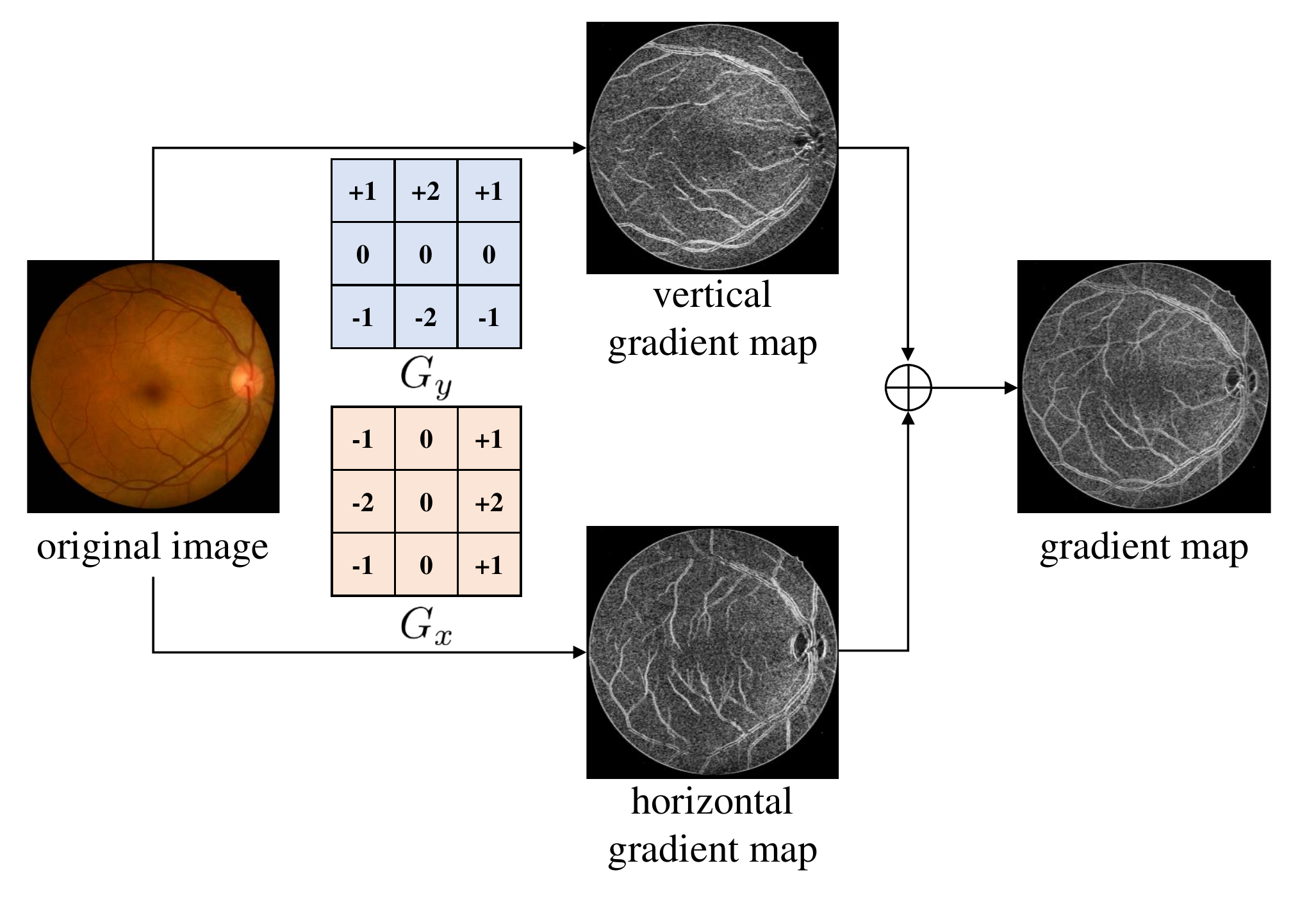}
     \caption{Generation of the gradient map. Please note the original image is the reconstruction objective of the intensity decoder and the gradient map is the reconstruction target of the edge decoder. Keys: $G_x$ - Sobel operator in the horizontal direction; $G_y$ - Sobel operator in the vertical direction.  }
     \label{stsk}
     \vspace{-10pt}
\end{figure}}
Compared with natural images, ophthalmic images are more similar across different samples, which makes diagnoses of eye diseases challenging. Some detailed information, such as that on the retinal vessels, is important for disease dignosis but is easily to be ignored due to the redundancy in image information. For example, the reconstructed images in \cite{bao2021beit} are blurry with little edge information, which is not suitable for medical images.
Therefore, we employ two decoders, namely an intensity decoder and an edge decoder, to encourage the network to learn representations containing both local and global information. The intensity decoder and edge decoder share the same network structure and the same input $\tilde{g}^d$. As shown in Fig. \ref{pipeline}, $\tilde{g}^d$ denotes the full set of patches consisting both the encoded visible patches $g^d$ and the masked patches. $\tilde{g}^d$ is simultaneously fed into the intensity decoder and the edge decoder. The difference between the two decoders lies in the reconstruction objectives.

As shown in Fig. \ref{stsk}, taking the fundus image as an example, the left side is the original input, which is the reconstruction objective of the intensity decoder. We apply Sobel operators \cite{kanopoulos1988design} at both horizontal $G_x$ and vertical $G_y$ directions to the original image, to get the horizontal gradient map and the vertical gradient map. Afterwards, we integrate these two gradient maps and obtain the gradient map of the fundus image, which is the reconstruction target of the edge decoder. We apply this operation to all 2D images and the 2D slices of each 3D volume. Compared with the original image, the gradient map uniformly characterizes the edge of the retinal structure and more clearly depicts tiny retinal vessels. In summary, in stage $P$, with trade-off parameters $\lambda_1$ and $\lambda_2$, the total objective function of our self-supervised learning framework is
{\setlength\abovedisplayskip{0.3cm}
\setlength\belowdisplayskip{0.2cm}
\begin{equation}
    \label{Lssl}
    \mathcal L_{ssl} = \lambda_1 \mathcal L_{i} +\lambda_2 \mathcal L_{e},
% \vspace{-0.2cm}
\end{equation}}

\noindent where $\mathcal L_{i}$ and $\mathcal L_{e}$ are mean squared error (MSE) losses of the masked patches between the predictions from the intensity/edge decoders and the corresponding targets. $\lambda_1$ and $\lambda_2$ are set as 0.5 and 0.5 to make the network concentrate equally on global intensity information and local edge information of the ophthalmic images of interest.
\vspace{-10pt}

\section{Experiments and Results}
\subsection{Experimental Setup}
In the pre-training phase, the input images of \emph{mmOphth}-v1 are downsampled as 224 $\times$ 224 for 2D images and 112 $\times$ 224 $\times$ 112 for 3D images. The batch size is 64 for 2D and 4 for 3D. The data augmentation strategy is a combination of random color jitter, random grayscaling, random cropping and random horizontal flipping. The model is optimized by an AdamW optimizer \cite{adamw} with an initial learning rate of 0.0005. Our model is implemented in PyTorch \cite{NEURIPS2019_9015} with 2 NVIDIA GeForce RTX 3090 GPUs, which takes 50 epochs and 20 hours to converge. In the fine-tuning phase, the input keeps consistent with the aforementioned settings. AdamW is also used as the optimizer with an initial learning rate of 0.0001 and the batch sizes are respectively set to be 8 and 1 for 2D and 3D images. Since all downstream tasks are classification tasks, we employ the area under curve (AUC), accuracy, precision, recall, F1-score and Kappa as our evaluation metrics. Details of the \emph{mmOphth}-v1 ophthalmic dataset and the evaluation datasets are presented in Fig. A1 and Table A1 of the appendix.
\vspace{-15pt}
{\setlength\abovedisplayskip{0.3cm}
\setlength\belowdisplayskip{0.3cm}
\begin{table}[htb]
\caption{Results obtained by fine-tuning on four 2D datasets. Rand denotes randomly-initialized model parameters. ViT-base and ViT-large respectively denote ViT-base-patch16-224 and ViT-large-patch16-224. - denotes the result is not available from the original article. (Unit: \%)}
\vspace{5pt}
\scalebox{0.7}{
\begin{tabular}{c|cccccccccccc}
\Xhline{1.2pt}
\multirow{3}{*}{Method}           & \multicolumn{6}{c}{Ichallenge-AMD}                                             & \multicolumn{6}{c}{Ichallenge-PM}                         \\ \cline{2-13}
& AUC   & Accuracy  & Precision & Recall & F1-score & \multicolumn{1}{c|}{Kappa} & AUC   & Accuracy  & Precision & Recall & F1-score & Kappa \\ \cline{2-13}
           &\multicolumn{12}{c}{Convolutional Neural Network}                                                                                          \\ \hline
Rand \cite{li2020selfsupervised} & 77.19 & 87.09     & 82.98     & 77.82  & 79.27    & \multicolumn{1}{c|}{-}     & 98.04 & 97.66     & 97.30     & 98.04  & 97.53    & -     \\ \
Invariant \cite{ye2019unsupervised}   & 81.62 & 87.51     & 81.92     & 81.62  & 81.35    & \multicolumn{1}{c|}{-}     & 98.02 & 97.84     & 97.56     & 98.02  & 97.75    & -     \\ \
Li et al. \cite{li2020selfsupervised}   & 83.17 & 89.37     & 85.71     & 83.17  & 83.67    & \multicolumn{1}{c|}{-}     & 98.41 & 98.38& 98.31     & 98.41  & 98.33    & -     \\  \hline
     Method      &\multicolumn{12}{c}{Vision Transformer}                                                                                                    \\ \hline
Rand        & 64.92 & 82.25  &     77.47      &      64.92  &    67.59      & \multicolumn{1}{c|}{36.78}      &    96.65   &     96.48      &    96.13       &  96.65      &     96.36     & 92.73      \\
ImageNet           & 73.75      &   86.00        &      82.68     &     73.75  &    76.79      & \multicolumn{1}{c|}{54.00}      &   97.70    &   97.74        &     97.60      &   97.70     &     97.65     &     95.30 \\  
SiT \cite{atito2021sit}         &   78.81    &   88.25    &      85.32     &     78.81   &       81.38   & \multicolumn{1}{c|}{62.92}      &  97.80     &    97.49       &    97.10       &    97.80    &     97.41     &   94.82    \\ 
Ours (ViT-base)             &  83.13    &   89.95   &    86.86       &    83.13    &    84.78       & \multicolumn{1}{c|}{69.60}      &   98.22    &    98.24     & 98.12  &    98.22    &     98.17     &    96.35   \\ \
\textbf{Ours (ViT-large)}         &  \textbf{85.85}    &  \textbf{90.45}    &   \textbf{86.44}         &    \textbf{85.85}    &      \textbf{86.14}      & \multicolumn{1}{c|}{\textbf{72.28}}      &    \textbf{98.53}   &    \textbf{98.24}       &   \textbf{97.90}&  \textbf{98.53}      &    \textbf{98.18}      &    \textbf{96.37}   \\ \Xhline{1.2pt}
\multirow{3}{*}{Method}            & \multicolumn{6}{c}{OCTA-500 (2D)}                                             & \multicolumn{6}{c}{GAMMA (2D)}                            \\ \cline{2-13}
 & AUC   & Accurancy & Precision & Recall & F1-score & \multicolumn{1}{c|}{Kappa} & AUC   & Accurancy & Precision & Recall & F1-score & Kappa \\ \cline{2-13}
          &\multicolumn{12}{c}{Vision Transformer}                                                                                                    \\ \hline
Rand     &73.63   &  73.60     &      74.35     &     73.63      &    73.41    &      \multicolumn{1}{c|}{47.24} &  90.04     &    90.00 & 90.04  &   90.04    & 90.00          & 78.01      \\  
ImageNet    &  74.65     &   74.60    &      76.50     &     74.65      &    74.16    &      \multicolumn{1}{c|}{49.25}     &   91.00    &     91.00      &     91.02      &  91.00      &    91.00      &  82.00     \\ \
SiT \cite{atito2021sit}         &  81.72     &      81.73     &     81.73      &   81.72     &       81.73   & \multicolumn{1}{c|}{63.45}      & 93.83      &  93.88         &   93.98        &   93.83     &    93.87      &   87.74    \\ \
Ours (ViT-base)    &    82.10     &   82.13    &   82.32      &     82.10      &   82.09    &      \multicolumn{1}{c|}{64.32}       & 94.92     &    94.90      &     94.90   &   94.92    &   94.90       &    89.80  \\ 
\textbf{Ours (ViT-large)}   &   \textbf{83.40}      &   \textbf{83.34}    &      \textbf{83.33}     &     \textbf{83.40}      &    \textbf{83.33}    &      \multicolumn{1}{c|}{\textbf{66.67}}       &   \textbf{97.00}    &      \textbf{97.00}     &     \textbf{97.02}      &    \textbf{97.00}    &    \textbf{97.00}      &   \textbf{94.00}    \\ \Xhline{1.2pt}
\end{tabular}\label{twoD}
}
% \vspace{-0.2cm}
\end{table}}
\vspace{-25pt}
\subsection{Comparison with State-of-the-art}

% For 2D This subsection finetunes our Uni4Eye under different pre-training strategies with different backbone networks like Convolutional Neural Networks (CNN) and Vision Transformers on four 2D ophthalmic datasets. As shown in Table. \ref{twoD}, when the network is trained from scratch, it will be the lowest performance in CNN or ViT on all the datasets. Although other state-of-the-art (SOTA) self-supervised methods like invariant \cite{ye2019unsupervised}, Li et al. \cite{li2020selfsupervised} based on CNN or SiT based on ViT could improve the performance of the network, it is worth emphasizing that our Uni4Eye significantly outperform all the ViT pre-training strategies. Besides, our Uni4Eye also achieve around 2.2$\%$ improvement on F1-score over Li et al. in Ichallenge-AMD dataset.
We compare Uni4Eye with other SOTA SSL methods employing convolutional neural network (CNN) or ViT as the backbone.
The binary classification results of different pre-training methods on four 2D datasets are shown in Table \ref{twoD}.  Li et al. \cite{li2020selfsupervised} feeds paired fundus and FFA data into a CNN for self-supervised contrastive learning, and achieves SOTA performance on Ichallenge-AMD and Ichallenge-PM datasets. Self-supervised Vision Transformer (SiT) \cite{atito2021sit} conducts image reconstruction, rotation prediction and contrastive learning tasks for pre-training, which outperforms randomly-weighted initialization and ImageNet pre-training. Although these SSL methods are beneficial in improving the classification performance, it is worth emphasizing that our Uni4Eye outperforms all compared methods regardless of the backbone. On the Ichallenge-AMD dataset, our method outperforms the second best method in terms of the F1-score by 2.2\%.
\vspace{-20pt}
\begin{table}[htb] 
\centering
\caption{Results obtained by fine-tuning on 3D OCT volumes from the GAMMA and OCTA-500 datasets. (Unit: \%)}
\vspace{5pt}
\scalebox{0.72}{
\begin{tabular}{c|cccccccccccc} 
\Xhline{1.2pt}
\multirow{3}{*}{Method}    & \multicolumn{6}{c}{GAMMA (3D)}                                            & \multicolumn{6}{c}{OCTA-500 (3D)}                          \\ \cline{2-13}
   & AUC & Accurancy & Precision & Recall & F1-score & \multicolumn{1}{c|}{Kappa} & AUC & Accurancy & Precision & Recall & F1-score & Kappa \\ \cline{2-13}
 &\multicolumn{12}{c}{Convolutional Neural Network}     \\ \hline

Med3D \cite{chen2019med3d} &  86.23   &     86.73     &     87.24    &  86.23    &      86.50    & \multicolumn{1}{c|}{73.07}      &   66.07  &       67.87    &     78.29      &    66.07    &     63.17     &   33.27   \\ \hline
 Method&\multicolumn{12}{c}{Vision Transformer}  \\ \hline

Rand &   85.79  &    85.71       &       85.91    &   85.79     &     85.71     & \multicolumn{1}{c|}{71.46}      &   60.33  &       60.64    &     77.15      &    60.33    &    53.25      &   20.78    \\ 
ImageNet    &   85.28  &     85.71      &     86.01      &    85.28    &      85.50    & \multicolumn{1}{c|}{71.04}      &   65.12  &       66.87    &  75.45         &     65.12   &   62.32       &  31.27     \\ 
\textbf{Ours}        &  \textbf{86.39}   &     \textbf{86.73}      &    \textbf{86.90}       &   \textbf{86.39}     &    \textbf{86.57}      & \multicolumn{1}{c|}{\textbf{73.16}}  &  \textbf{66.18}   &    \textbf{67.87}       &      \textbf{76.13}     &    \textbf{66.18}    &     \textbf{63.75}     &     \textbf{33.43}                                                                                                \\ \Xhline{1.2pt}
\end{tabular}\label{3D}}
\vspace{-5pt}
\end{table}
% \vspace{-5pt}

For 3D downstream tasks, we fine-tune Uni4Eye on the OCT volumes from the GAMMA dataset and the OCTA-500 dataset. As shown in Table \ref{3D}, our proposed Uni4Eye performs better than random initialization and ImageNet pre-training. Please note that ImageNet pre-training means we only replace the patch embedding module of ViT with a 3D version, and maintain all other pre-trained parameters of ViT. Since there is relatively few amount of 3D ophthalmic data, the classification performance of the 3D model is worse than that of the 2D model.

\vspace{-15pt}
\begin{table}[]
\centering
\caption{Results obtained by first training a self-supervised model on \emph{mmOphth}-v1 with different mask ratios $\alpha$ and then fine-tuning on the Ichallenge-AMD dataset. (Unit: \%)}
\vspace{5pt}
\begin{tabular}{c|cccccc}
\Xhline{1.2pt}
Metrics         & AUC & Accurancy & Precision & Recall & F1-score & Kappa \\ \hline
$\alpha$=0.25 &  80.80   &     89.45      &    87.32       &   80.8     &      83.42    &  66.97     \\ 
$\alpha$=0.5  &   \textbf{85.85}    &  90.45    &   86.44       &    \textbf{85.85}    &      \textbf{86.14}    &  \textbf{72.28}     \\ 
$\alpha$=0.75 & 83.38    &     \textbf{91.00}      &    \textbf{89.56}       & 83.38       &   85.94  &  71.96      \\ \Xhline{1.2pt}
\end{tabular}\label{ratio}
\vspace{-10pt}
\end{table}
\vspace{-18pt}
\subsection{Reconstruction Results}
\begin{figure}[htb]
\centering
     \includegraphics[width=0.75\textwidth]{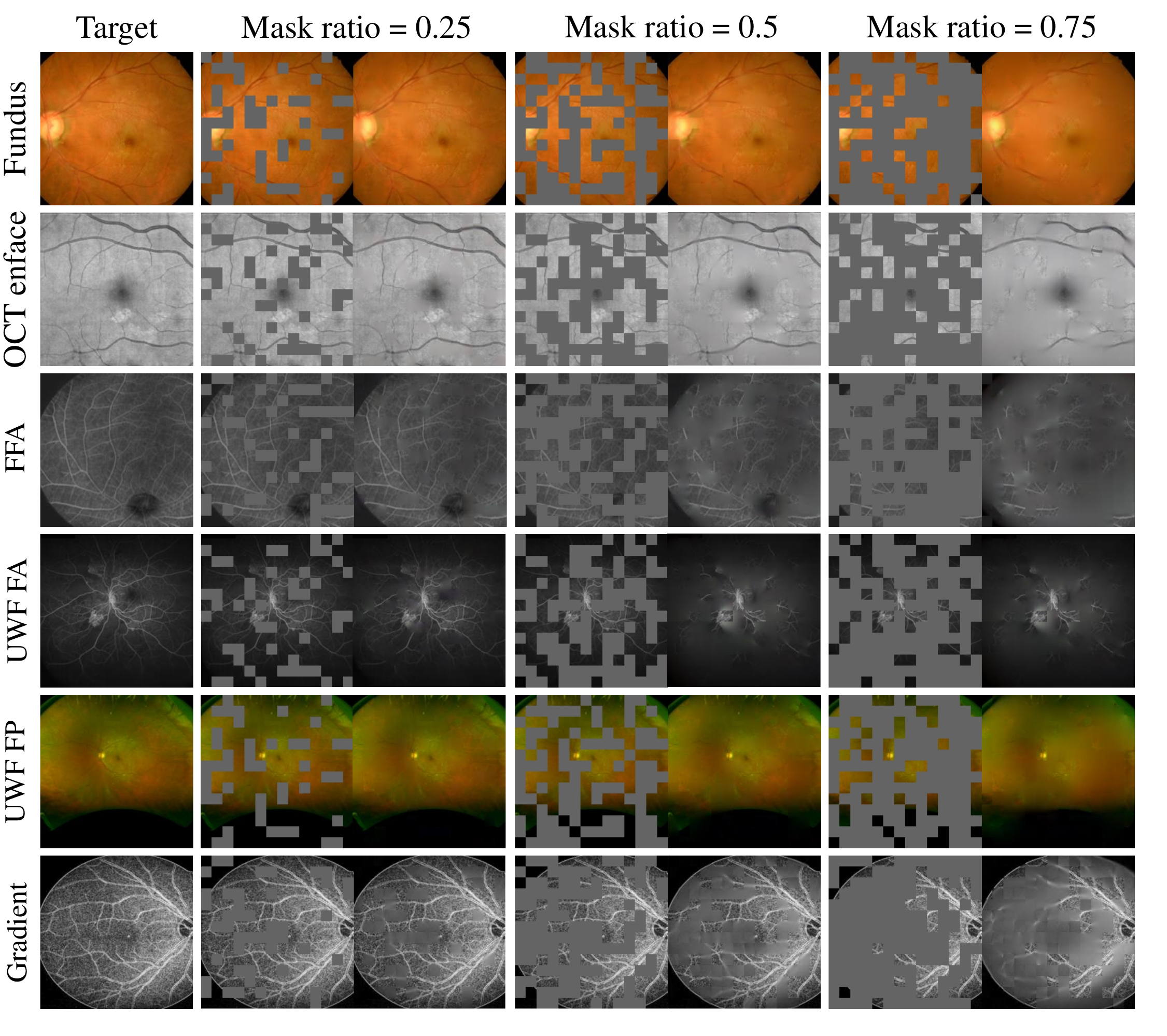}
     \caption{The reconstruction results for six common modalities (from top to bottom), with different mask ratios in stage $P$.}
     \label{mr}
\vspace{-12pt}
\end{figure}

We visualize the reconstruction results of different ophthalmic modalities from the same network pre-trained on \emph{mmOphth}-v1 to highlight the universality of our learned features. As shown in Fig. \ref{mr}, we feed the input of different modalities to the network and obtain the reconstruction results. We set the mask ratio in UPE as 25$\%$, 50$\%$, 75$\%$. It is clear that a smaller mask ratio enables the model to generate better reconstruction results. However, better reconstruction is not equivalent to better performance on downstream tasks. We fine-tune these three models on Ichallenge-AMD with the same settings. As shown in Table \ref{ratio}, the network pre-trained with a 50$\%$ mask ratio achieves the best performance on the specific downstream task of interest. For ophthalmic image analysis, this result may suggest the encoder cannot generate discriminative representations through a too-easy (mask ratio = 25\%) or a too-difficult (mask ratio = 75\%) reconstruction task. Ablation analysis results are presented in Tables A2-A3 of the appendix, demonstrating the importance of resolving the dimension barrier and that of employing the dual-branch decoder.
\vspace{-5pt}
\section{Conclusion}
\vspace{-3pt}
This paper proposes a simple, unified and powerful self-supervised framework, namely Uni4Eye, for ophthalmic image analysis. Specifically, by modifying the patch embedding module to generate UPE in ViT, Uni4Eye can easily break the dimension barrier and process both 2D and 3D images. We also design a dual-decoder structure based on the MIM task, to make Uni4Eye take advantage of not only intensity information but also edge information in ophthalmic images. Extensive experiments on four 2D datasets and two 3D datasets show that our Uni4Eye achieves better classification performance than representative SOTA methods for eye disease diagnoses. Our results also demonstrate the potential of MIM for self-supervised pre-training in various medical image analyses. Our future work will involve investigating the feasibility of our framework for other types of medical images and exploring methods to further improve the efficiency of our framework.
\clearpage
\appendix
\section{Appendix}
\renewcommand\thetable{\Alph{section}\arabic{table}}
\setcounter{table}{0}
\renewcommand{\thetable}{A\arabic{table}}
\renewcommand\thefigure{\Alph{section}\arabic{figure}}  
\setcounter{figure}{0}
\renewcommand{\thefigure}{A\arabic{figure}}
\vspace*{\fill}
\begin{figure}[htb]
\centering
     \includegraphics[width=\textwidth]{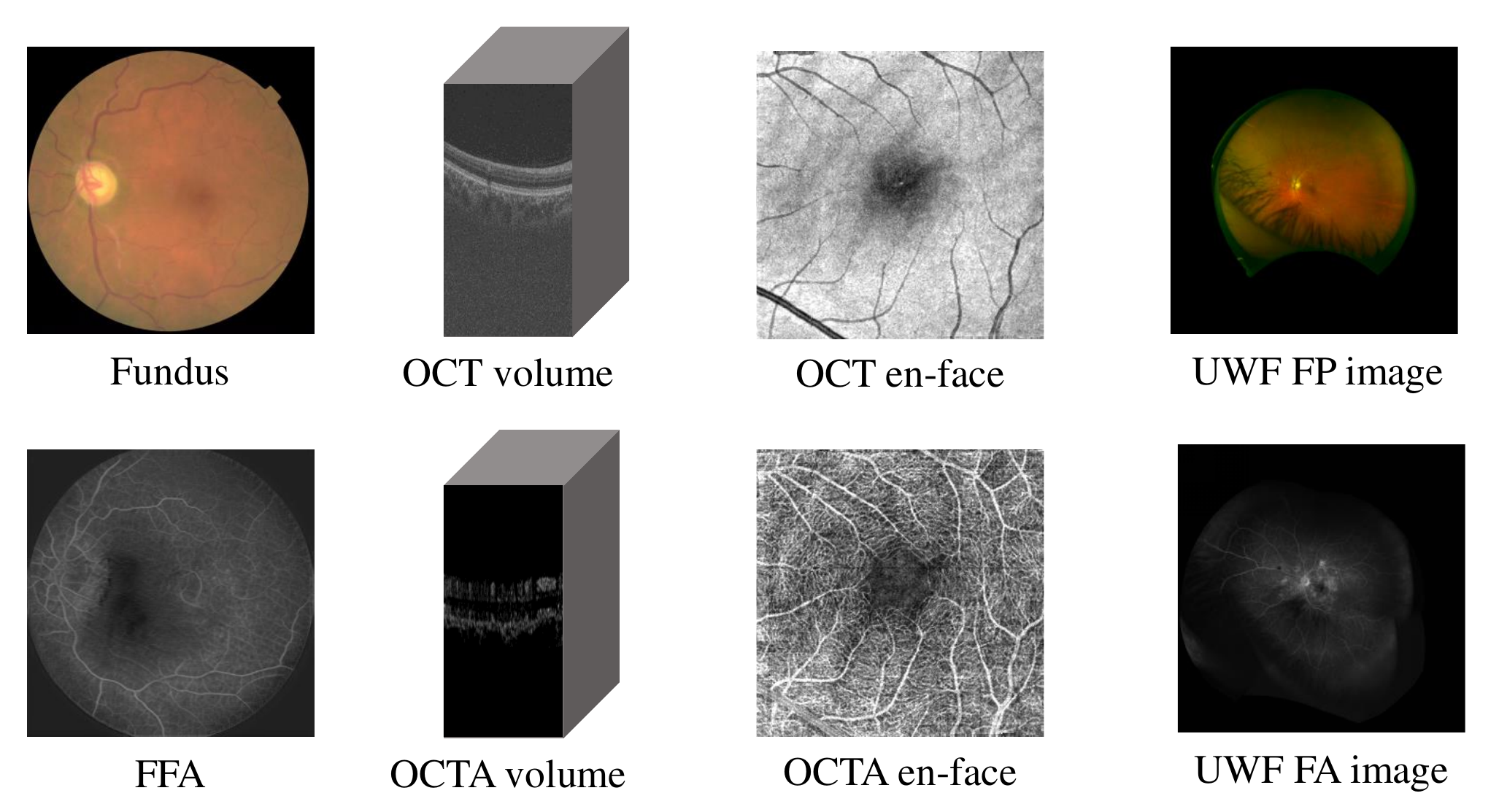}
     \caption{Demonstration of the \emph{mmOphth}-v1 dataset for self-supervised training.}
     \label{dataset}
\end{figure}
\vspace*{\fill}
\clearpage
\vspace*{\fill}
\begin{table}[htbp]
\caption{Details of the \emph{mmOphth}-v1 ophthalmic dataset for self-supervised pre-training and evaluation datasets for downstream classification tasks. Keys: UWF FA - ultra-widefield fluorescein angiography, UWF FP - ultra-widefield fundus photography, FFA - Fundus Fluorescein Angiography.}
\vspace{5pt}
\begin{tabular}{ccc}
\Xhline{1.2pt}
\multicolumn{3}{c}{\emph{mmOphth}-v1 Ophthalmic Dataset}                                                                                       \\ \Xhline{1.2pt}
\multicolumn{1}{c|}{Name}            & \multicolumn{1}{c|}{Modality}                           & Sample Size                    \\ \hline
\multicolumn{1}{c|}{OCTA-500}        & \multicolumn{1}{c|}{OCT, OCTA, OCT en-face, OCTA en-face} & 500, 500, 3,000, 3,000 \\ 
\multicolumn{1}{c|}{GAMMA}           & \multicolumn{1}{c|}{Fundus, OCT}                          & 100, 100    \\ 
\multicolumn{1}{c|}{EyePACS}         & \multicolumn{1}{c|}{Fundus}                               & 88,702                 \\ 
\multicolumn{1}{c|}{PRIME-FP20 }      & \multicolumn{1}{c|}{UWF FP, UWF FA}                       & 15                     \\ 
\multicolumn{1}{c|}{Synthesized FFA} & \multicolumn{1}{c|}{FFA}                                  & 400                    \\ \Xhline{1.2pt}
\multicolumn{3}{c}{Downstream Task Evaluation Datasets}                                                                              \\ \Xhline{1.2pt}
\multicolumn{1}{c|}{Name}            & \multicolumn{1}{c|}{Modality}                           & Sample Size                    \\ \hline
\multicolumn{1}{c|}{OCTA-500}        & \multicolumn{1}{c|}{OCT, OCTA, OCT en-face, OCTA en-face} & 500, 500, 3,000, 3,000 \\ 
\multicolumn{1}{c|}{GAMMA}           & \multicolumn{1}{c|}{Fundus, OCT}                          & 100, 100    \\ 
\multicolumn{1}{c|}{Ichallenge-AMD}  & \multicolumn{1}{c|}{Fundus}                               & 400                    \\
\multicolumn{1}{c|}{Ichallenge-PM}   & \multicolumn{1}{c|}{Fundus}                               & 400                    \\ \Xhline{1.2pt}
\end{tabular} \label{datasets}
\end{table}

\begin{table}[]
\centering
\vspace{-30pt}
\caption{Ablation analysis results on the GAMMA dataset. Keys: 2D - Pre-training with 2D data; 3D - Pre-training with 3D data. (Unit: \%)}
\vspace{5pt}
\begin{tabular}{@{}cccccccc@{}}
\Xhline{1.2pt}
2D & 3D    & AUC& Acc & Precision &Recall &F1-score&Kappa \\ \Xhline{1.2pt}
\checkmark& &        89.52  & 90.82  & 90.09  & 89.52  & 89.80 & 79.59        \\ 
  &\checkmark&       88.74 & 89.80           & 88.74 & 88.74  & 88.74  &    77.48   \\
  \checkmark  &  \checkmark& \textbf{92.65}    &  \textbf{94.90}    &   \textbf{96.38}         &    \textbf{92.65}    &      \textbf{94.15}      & \textbf{88.34}    \\ \Xhline{1.2pt}
% 2\cite{10.1007/978-3-030-87000-3_15} & \Checkmark& &                 &    x        & 0.6730        \\     \\ \bottomrule
\end{tabular}
\label{ablation1}
\vspace{-15pt}
\end{table}

\begin{table}[]
\centering
\vspace{-20pt}
\caption{Ablation analysis results of the intensity reconstruction decoder and the edge reconstruction decoder on the Ichallenge-AMD dataset. (Unit: \%)}
\vspace{5pt}
\begin{tabular}{@{}cccccccc@{}}
\Xhline{1.2pt}
Intensity & Edge    & AUC& Acc & Precision &Recall &F1-score&Kappa \\ \Xhline{1.2pt}
\checkmark& &        84.64  & 89.20  & 84.97  & 84.64  & 84.50 & 69.01        \\ 
  &\checkmark&       84.80 & 89.45           & 84.80 & 84.80  & 84.80  &    69.61   \\
  \checkmark  &  \checkmark& \textbf{85.85}    &  \textbf{90.45}    &   \textbf{86.44}         &    \textbf{85.85}    &      \textbf{86.14}      & \textbf{72.28}    \\ \Xhline{1.2pt}
% 2\cite{10.1007/978-3-030-87000-3_15} & \Checkmark& &                 &    x        & 0.6730        \\     \\ \bottomrule
\end{tabular}
\label{ablation}
\end{table}
\vspace*{\fill}
\clearpage
\bibliographystyle{splncs04}
\bibliography{ref1}

\begin{thebibliography}{10}
\providecommand{\url}[1]{\texttt{#1}}
\providecommand{\urlprefix}{URL }
\providecommand{\doi}[1]{https://doi.org/#1}

\bibitem{atito2021sit}
Atito, S., Awais, M., Kittler, J.: Sit: Self-supervised vision transformer.
  arXiv preprint \textcolor{blue}{arXiv: 2104.03602}  (2021)

\bibitem{bao2021beit}
Bao, H., Dong, L., et~al.: Beit: Bert pre-training of image transformers. In:
  International Conference on Learning Representations, ICLR (2022)

\bibitem{cai2022corolla}
Cai, Z., Lin, L., He, H., Tang, X.: Corolla: An efficient multi-modality fusion
  framework with supervised contrastive learning for glaucoma grading. arXiv
  preprint \textcolor{blue}{arXiv: 2201.03795}  (2022)

\bibitem{chaitanya2020contrastive}
Chaitanya, K., Erdil, E., Karani, N., Konukoglu, E.: Contrastive learning of
  global and local features for medical image segmentation with limited
  annotations. In: {Advances in Neural Information Processing Systems,
  NeurIPS}. vol.~33 (2020)

\bibitem{CHEN2019101539}
Chen, L., Bentley, P., et~al.: Self-supervised learning for medical image
  analysis using image context restoration. IEEE Transactions on Medical
  Imaging  \textbf{58},  101539 (2019). \doi{10.1016/j.media.2019.101539.}

\bibitem{chen2019med3d}
Chen, S., Ma, K., et~al.: Med3d: Transfer learning for 3d medical image
  analysis. arXiv preprint \textcolor{blue}{arXiv: 1904.00625}  (2019)

\bibitem{chen2020simple}
Chen, T., Kornblith, S., Norouzi, M., Hinton, G.: A simple framework for
  contrastive learning of visual representations. arXiv preprint
  \textcolor{blue}{arXiv: 2002.05709}  (2020)

\bibitem{cordeiro2021longremix}
Cordeiro, F.R., Sachdeva, R., et~al.: Longremix: Robust learning with high
  confidence samples in a noisy label environment. arXiv preprint
  \textcolor{blue}{arXiv: 2103.04173}  (2021)

\bibitem{deng2009imagenet}
Deng, J., Dong, W., Socher, R., Li, L.J., Li, K., Fei-Fei, L.: Imagenet: A
  large-scale hierarchical image database. In: Proceedings of the IEEE/CVF
  Conference on Computer Vision and Pattern Recognition, CVPR. pp. 248--255
  (2009)

\bibitem{donahue2019large}
Donahue, J., Simonyan, K.: Large scale adversarial representation learning. In:
  Advances in Neural Information Processing Systems, NeurIPS. vol.~32 (2019)

\bibitem{dosovitskiy2021image}
Dosovitskiy, A., Beyer, L., et~al.: An image is worth 16x16 words: Transformers
  for image recognition at scale. arXiv preprint \textcolor{blue}{arXiv:
  2010.11929}  (2021)

\bibitem{gidaris2018unsupervised}
Gidaris, S., Singh, P., Komodakis, N.: Unsupervised representation learning by
  predicting image rotations. arXiv preprint \textcolor{blue}{arXiv:1803.07728}
   (2018)

\bibitem{goodfellow2014generative}
Goodfellow, I.J., Pouget-Abadie, J., et~al.: Generative adversarial nets. In:
  Advances in Neural Information Processing Systems, NeurIPS. vol.~27 (2014)

\bibitem{JOINED}
He, H., Lin, L., Cai, Z., Tang, X.: Joined : Prior guided multi-task learning
  for joint optic disc/cup segmentation and fovea detection. In: International
  Conference on Medical Imaging with Deep Learning, MIDL (2022)

\bibitem{mae}
He, K., Chen, X., Xie, S., Li, Y., Dollár, P., Girshick, R.: Masked
  autoencoders are scalable vision learners. arXiv preprint
  \textcolor{blue}{arXiv: 2111.06377}  (2021)

\bibitem{he2020momentum}
He, K., Fan, H., Wu, Y., Xie, S., Girshick, R.: Momentum contrast for
  unsupervised visual representation learning. In: Proceedings of the IEEE/CVF
  Conference on Computer Vision and Pattern Recognition, CVPR. pp. 9729--9738
  (2020)

\bibitem{huang2021lesionbased}
Huang, Y., Lin, L., et~al.: Lesion-based contrastive learning for diabetic
  retinopathy grading from fundus images. In: International Conference on
  Medical Image Computing and Computer Assisted Intervention, MICCAI 2021. vol.
  12902, pp. 113--123. Springer, Cham (2021).
  \doi{10.1007/978-3-030-87196-3_11}

\bibitem{kanopoulos1988design}
Kanopoulos, N., Vasanthavada, N., Baker, R.L.: Design of an image edge
  detection filter using the sobel operator. {IEEE Journal of Solid-state
  Circuits}  \textbf{23}(2),  358--367 (1988)

\bibitem{li2021rotation}
Li, X., Hu, X., et~al.: Rotation-oriented collaborative self-supervised
  learning for retinal disease diagnosis. IEEE Transactions on Medical Imaging
  \textbf{40}(9),  2284--2294 (2021)

\bibitem{li2020selfsupervised}
Li, X., Jia, M., Islam, M.T., Yu, L., Xing, L.: Self-supervised feature
  learning via exploiting multi-modal data for retinal disease diagnosis. IEEE
  Transactions on Medical Imaging  \textbf{39}(12),  4023--4033 (2020)

\bibitem{lin2020sustech}
Lin, L., Li, M., Huang, Y., Cheng, P., Xia, H., Wang, K., Yuan, J., Tang, X.:
  The sustech-sysu dataset for automated exudate detection and diabetic
  retinopathy grading. Scientific Data  \textbf{7}(1),  1--10 (2020)

\bibitem{lin2021bsdanet}
Lin, L., Wang, Z., et~al.: Bsda-net: A boundary shape and distance aware joint
  learning framework for segmenting and classifying octa images. In:
  International Conference on Medical Image Computing and Computer Assisted
  Intervention, MICCAI 2021. vol. 12098, pp. 65--75. Springer, Cham (2021).
  \doi{10.1007/978-3-030-87237-3_7}

\bibitem{adamw}
Loshchilov, I., Hutter, F.: Fixing weight decay regularization in adam. arXiv
  preprint \textcolor{blue}{arXiv: 1711.05101}  (2017)

\bibitem{oliver2019realistic}
Oliver, A., Odena, A., Raffel, C., Cubuk, E.D., Goodfellow, I.J.: Realistic
  evaluation of deep semi-supervised learning algorithms. In: Advances in
  Neural Information Processing Systems, NeurIPS. vol.~31 (2019)

\bibitem{NEURIPS2019_9015}
Paszke, A., Gross, S., et~al.: Pytorch: An imperative style, high-performance
  deep learning library. In: Advances in Neural Information Processing Systems,
  NeurIPS, vol.~32 (2019)

\bibitem{taleb20203d}
Taleb, A., Loetzsch, W., et~al.: 3d self-supervised methods for medical
  imaging. In: Advances in Neural Information Processing Systems, NeurIPS.
  vol.~33 (2020)

\bibitem{tan2020efficientnet}
Tan, M., Le, Q.V.: Efficientnet: Rethinking model scaling for convolutional
  neural networks. In: International Conference on Machine Learning, ICML. pp.
  6105--6114 (2019)

\bibitem{maskfeat}
Wei, C., Fan, H., Xie, S., Wu, C.Y., Yuille, A., Feichtenhofer, C.: Masked
  feature prediction for self-supervised visual pre-training. arXiv preprint
  \textcolor{blue}{arXiv: 2112.09133}  (2021)

\bibitem{ye2019unsupervised}
Ye, M., Zhang, X., Yuen, P.C., Chang, S.F.: Unsupervised embedding learning via
  invariant and spreading instance feature. In: Proceedings of the IEEE/CVF
  Conference on Computer Vision and Pattern Recognition, CVPR. pp. 6210--6219
  (2019)

\bibitem{zhou2021preservational}
Zhou, H.Y., Lu, C., et~al.: Preservational learning improves self-supervised
  medical image models by reconstructing diverse contexts. In: The IEEE
  International Conference on Computer Vision, ICCV. pp. 3499--3509 (2021)

\bibitem{Cube}
Zhuang, X., Li, Y., Hu, Y., Ma, K., Yang, Y., Zheng, Y.: Self-supervised
  feature learning for 3d medical images by playing a rubik's cube. In:
  International Conference on Medical Image Computing and Computer Assisted
  Intervention, MICCAI 2019. vol. 11767, pp. 420--428. Springer, Cham (2019).
  \doi{10.1007/978-3-030-32251-9_46}

\end{thebibliography}
\end{document}